\documentclass[english]{article}
\usepackage[T1]{fontenc}
\usepackage[latin9]{inputenc}
\usepackage{geometry}
\usepackage{xcolor}
\usepackage{enumitem}
\usepackage{amsmath}
\usepackage{amsthm}
\usepackage{amssymb}
\usepackage{graphicx}
\usepackage[authoryear]{natbib}

\makeatletter

\providecommand{\tabularnewline}{\\}

\theoremstyle{plain}
\newtheorem{thm}{\protect\theoremname}

\usepackage{hyperref}
\usepackage{authblk}

\newtheorem{algorithm}{Algorithm}
\newtheorem{assumption}{Assumption}

\makeatother

\usepackage{babel}
\providecommand{\theoremname}{Theorem}

\begin{document}
\title{Online Control of the False Discovery Rate under ``Decision Deadlines''}
\author{Aaron Fisher\thanks{Foundation Medicine Inc.; 150 Second St, Cambridge, MA 02141. \texttt{afishe27@alumni.jh.edu}}}
\maketitle
\begin{abstract}
Online testing procedures aim to control the extent of false discoveries
over a sequence of hypothesis tests, allowing for the possibility
that early-stage test results influence the choice of hypotheses to
be tested in later stages. Typically, online methods assume that a
permanent decision regarding the current test (reject or not reject)
must be made before advancing to the next test. We instead assume
that each hypothesis requires an immediate \emph{preliminary }decision,
but also allows us to update that decision until a preset deadline.
Roughly speaking, this lets us apply a Benjamini-Hochberg-type procedure
over a moving window of hypotheses, where the threshold parameters
for upcoming tests can be determined based on preliminary results.
Our method controls the false discovery rate (FDR) at every stage
of testing, as well as at adaptively chosen stopping times. These
results apply even under arbitrary p-value dependency structures.
\end{abstract}
\textbf{\medskip{}
}


\section{INTRODUCTION}

Scientific discoveries form an ongoing, ever-evolving process. Each
new experiment offers an opportunity to suggest new hypotheses based
on results that have come before. Traditionally, the hypotheses researchers
plan to test in an experiment are prespecified before any data from
the experiment is visible, as this facilitates control of either the
false discovery rate (FDR; \citealp{Benjamini1995-sz}) or the probability
of producing \emph{any} false positives (the familywise error rate,
or FWER; see, for example \citealp{Efron2016-bj}) within that experiment. 

In contrast to fully prespecified procedures, \emph{online} procedures
test hypotheses sequentially, and allow the results of preliminary
tests to inform choices about which hypotheses to focus on in future
tests \citep{Foster2008-ta}. These procedures typically require
that error rates be controlled at every stage of the sequence (e.g.,
\citealp{Javanmard2015-tn,Ramdas2017-rm}). The online setting is
increasingly relevant to large-scale experimentation, and to repeated
analyses of public datasets \citep{Aharoni2014-lv}. At a high level,
online testing can be seen as an abstraction of the scientific process
itself \citep{Xu2020-qx}.

Online testing problems also arise when users must quickly decide
how to take action in response to a stream of data. Applications range
from monitoring credit card transactions for instances of fraud \citep{Zrnic2020-xl}
to deciding how to assign treatments to sequences of patients. Here,
hypotheses quickly become irrelevant, and so final decisions must
be made without delay. In other words, a discovery has little value
if the opportunity to act on it has passed.

On the other hand, streams of hypothesis tests do not always require
immediate, permanent decisions. In particular, if our goal is to maintain
a growing library of scientific knowledge (\citealp{Aharoni2014-lv}),
then hypotheses can remain relevant long after they are tested. Here,
discoveries remain valuable even if they are made \emph{retroactively}.
\textcolor{lightgray}{}

With this mind, we study scenarios where limited forms of decision
updating still add value. Specifically, we consider the setting where
each hypothesis requires an immediate, preliminary decision (reject
or not reject), but also allows us to update that decision until some
preset deadline. To incorporate these ``decision deadlines,'' we
blend two existing procedures: the well-known, offline \citeauthor{Benjamini1995-sz}
(BH, \citeyear{Benjamini1995-sz}) procedure, and an online procedure
known as \emph{significance levels based on number of discoveries
}(LOND; \citealt{Javanmard2015-tn}). Our procedure can reduce to
LOND if all decisions must be made immediately, or to BH if all decisions
can be updated indefinitely. Because the option for decision updates
is limited to evolving subset of ``active'' hypotheses, we refer
to our approach as \emph{significance thresholds based on active discoveries}
(TOAD).

We show that our approach provides online FDR control under arbitrary
p-value dependency structures. We also allow the parameters used in
setting significance thresholds to be determined based on preliminary
results, which, in turn, lets us control FDR at adaptively determined
stopping times. That is, we can still control FDR even if analysts
end their experiments early due especially strong preliminary results. 

\subsection{Outline}

The remained of our paper is organized as follows. Section \ref{sec:RELATED-LITERATURE}
discusses the advantages of our approach relative to other methods
in the literature. Section \ref{sec:Notation} introduces relevant
notation. Section \ref{sec:Thresholds-Based-on} presents the TOAD
procedure along with its FDR guarantees. Section \ref{sec:Simulations}
uses simulations to compare the power of TOAD to the power of similar
methods introduced by \citet{Zrnic2020-xl}. We conclude with a discussion
of several extensions and possible future directions (namely, adaptive
hypothesis reordering, and incorporating the concept of ``decaying
memory''). All proofs are provided in the supplementary materials.
These proofs use a combination of methods from \citealp{Blanchard2008-qc,Javanmard2015-tn,Ramdas2017-rm};
and \citealp{Zrnic2021-jj}.

\subsection{\label{sec:RELATED-LITERATURE}Related Literature}

In recent work that most closely resembles our own, \citet{Zrnic2020-xl}
propose two online methods for applying Benjamini-Hochberg procedures
to \emph{batches} of hypotheses (referred to as $\text{Batch}_{\text{BH}}$
and $\text{Batch}_{\text{BH}}^{\text{PRDS}}$). This batch testing
framework forms a special case of online testing under decision deadlines,
where the deadline for each test in a batch is the time of the last
test in that batch. 

Our work differs from that of \citet{Zrnic2020-xl} in three substantial
ways. First, our framing in terms of ``deadlines'' is more flexible
than the batch structure used by \citet{Zrnic2020-xl}. \textcolor{black}{Second,
we will show analytically that TOAD is at least as powerful as $\text{Batch}_{\text{BH}}^{\text{PRDS}}$,
and will show in simulations that it is typically more powerful (see
Sections \ref{sec:Thresholds-Based-on} \& \ref{sec:Simulations},
as well as the supplementary materials).} Finally, we prove FDR control
under arbitrary p-value dependencies, whereas \citet{Zrnic2020-xl}
prove FDR control under an assumption of independence across batches\textcolor{black}{}.

In another approach that is conceptually similar to ours, \citet{Zrnic2021-jj}
suggest ``revisiting'' hypotheses by allowing duplicated test statistics
in later stages (see their Section 3). We differ from \citet{Zrnic2021-jj}
in that we simultaneously update all active hypotheses at every stage
rather than updating hypotheses individually. 

The fact that TOAD provides online FDR control under arbitrary p-value
dependencies is nontrivial in the literature. To our knowledge, there
is only one other existing online method that controls the FDR under
arbitrary dependencies without imposing other restrictions \citep{Xu2020-qx}.
Typically, online bounds on the FDR require an independence condition
on the p-values \citep{Ramdas2017-rm,Ramdas2018-qu,Tian2019-fl,Zrnic2020-xl,Zrnic2021-jj}.
Alternatively, many existing methods focus on controlling either the
``modified'' FDR or the marginal FDR (\citealt{Foster2008-ta,Aharoni2014-lv,Ramdas2017-rm,Ramdas2018-qu,Tian2019-fl,Zrnic2021-jj}),
rather than the traditional FDR (\citealt{Benjamini1995-sz}). Some
additional online methods do control FDR without an independence assumption
(\citealt{Javanmard2015-tn,Javanmard2018-rz}; see also \citealp{Zrnic2021-jj}).
However, unlike TOAD, these methods do not allow the user to selectively
ignore future hypotheses based on preliminary results (see discussion
in Section \ref{subsec:Examples}, below).

\subsection{Notation\label{sec:Notation}}

Let $H_{1},H_{2}\dots$ be a possibly infinite sequence of hypotheses,
and let $P_{1},P_{2},\dots$ be p-values associated with each hypothesis.
Such a sequence can result either from a growing (streaming) dataset
with an increasing number of subgroups, or from a series of distinct
questions applied to a fixed dataset. As we will see in Section \ref{subsec:Examples},
many forms of online decision making can be captured by this framework. 

We consider the setting where, at each stage $t$ of testing, we observe
the next p-value $P_{t}$ and must make an immediate, preliminary
decision to reject or not reject $H_{t}$. However, we are also permitted
to update our decision up until a preset\emph{ }deadline $d_{t}\geq t$
(i.e., the decision for $H_{t}$ cannot be altered after stage $d_{t}$).
We use $\mathcal{C}_{t}$ to denote the set of ``active'' candidate
hypotheses for which decisions can still be updated at stage $t$,
i.e., $\mathcal{C}_{t}=\{i\leq t:d_{i}\geq t\}$. For example, if
we allow rejection decisions to be updated indefinitely, then $d_{t}=\infty$
and $\mathcal{C}_{t}=\{1,\dots,t\}$ for all $t$. If we require final
decisions instantaneously, then $\mathcal{C}_{t}=\{d_{t}\}=\{t\}$. 

Let $\mathcal{R}_{t}\subseteq\{1,\dots,t\}$ denote the indices for
the hypotheses that we reject at stage $t$. Again, any differences
in the sets of hypotheses rejected at consecutive stages must be limited
to the hypotheses whose deadlines have not yet passed (i.e., $\{\mathcal{R}_{t}\setminus\mathcal{C}_{t}\}=\{\mathcal{R}_{t-1}\setminus\mathcal{C}_{t}\}$).

We define $\mathcal{H}_{0}\subseteq\mathbb{N}$ to be the indices
corresponding to true null hypotheses, and define the FDR at time
$t$ to be 
\[
\text{FDR}(t)=\mathbb{E}\left[\frac{\vert\mathcal{H}_{0}\cap\mathcal{R}_{t}\vert}{1\vee\vert\mathcal{R}_{t}\vert}\right],
\]
where $a\vee b$ denotes the maximum over $\{a,b\}$. We use $\alpha$
to denote a desired level at which to control $\text{FDR}(t)$.

\section{Thresholds Based on Active Discoveries (TOAD)\label{sec:Thresholds-Based-on}}

We first describe the original LOND procedure \citep{Javanmard2015-tn},
as this method forms the original inspiration for our proposed method.
As input, LOND requires a sequence of nonnegative tuning parameters
$a_{1},a_{2},\dots$ satisfying $\sum_{i=1}^{\infty}a_{i}=1$. At
each stage $t$, LOND rejects $H_{t}$ if 
\begin{equation}
P_{t}\leq(\vert\mathcal{R}_{t-1}\vert+1)a_{t}\alpha.\label{eq:lond}
\end{equation}
Once a hypothesis is rejected, it remains rejected in all future stages.
\citet{Javanmard2015-tn} show that, under a condition on the joint
distribution of p-values, LOND controls FDR at every stage.

Building on this method, \citet{Zrnic2021-jj} propose a ``reshaped''
version of LOND that controls FDR under any p-value dependency structure
(see also Theorem 2.7 of \citealp{Javanmard2015-tn}). This version
additionally takes as input a sequence of so-called \emph{shape functions}
$\{\beta_{i}\}_{i=1}^{\infty}$. Following \citet{Blanchard2008-qc},
we say that $\beta$ is a shape function if there exists a probability
distribution $\nu$ on $\mathbb{R}_{>0}$ such that
\begin{equation}
\beta(r)=\mathbb{E}_{X\sim\nu}\left[X\times1(X\leq r)\right].\label{eq:shape}
\end{equation}
For example, when the number of stages $(t_{\text{max}})$ is finite,
\citeauthor{Blanchard2008-qc} consider setting $\nu$ to be the distribution
satisfying $\mathbb{P}_{X\sim\nu}(X=x)\propto1/x$ for each $x\in\{1,\dots,t_{\text{max}}\}$.
This produces the shape function $\beta(r)=r\left(\sum_{i'=1}^{t_{\text{max}}}1/i'\right)^{-1},$
which mimics the transformation employed by \citet{Benjamini2001-ps}.
To incorporate these shape functions $\{\beta_{i}\}_{i=1}^{\infty}$,
\citeauthor{Zrnic2021-jj} define the reshaped version of LOND to
reject each $H_{t}$ whenever $P_{t}\leq\beta_{t}(\vert\mathcal{R}_{t-1}\vert\vee1)a_{t}\alpha$.

Our proposed procedure differs from (reshaped) LOND in three key ways.
The first is a restriction, which is that we require users to select
a common function $\beta$ to be used at all stages. More specifically,
users can set $\beta$ to be either the identity function or a shape
function. Setting $\beta$ to be the identity function is the simplest
and most powerful option, but setting $\beta$ to be a shape function
will improve our FDR guarantee (see details in Section \ref{subsec:FDR-control}). 

The second two differences are expansions. Rather than prespecifying
all parameters $\{a_{i}\}_{i=1}^{\infty}$, we replace them with random
nonnegative random variables $\{A_{i}\}_{i=1}^{\infty}$ satisfying
$\sum_{i=1}^{\infty}A_{i}=1$. Of these, only $A_{1}$ must be specified
a priori. For the remaining test indices $i>1$, we define $\tau_{i}\leq i-1$
to be the stage by which the $i^{th}$ parameter $A_{i}$ must be
selected. That is, we require $A_{i}$ to be a deterministic function
of the first $\tau_{i}$ p-values $\{P_{i'}\}_{i'\leq\tau_{i}}$.
Setting $\tau_{i}=i-1$ is the simplest option, but we will see in
the next section setting $\tau_{i}<i-1$ can facilitate FDR control
when the test statistics are correlated (see also \citealp{Zrnic2021-jj}). 

We also expand on LOND by allowing users to update rejection decisions
for hypotheses whose deadlines have not yet passed. At each stage
$t$, our goal will be to find the \emph{largest} \emph{set} of rejected
indices $\mathcal{R}_{t}\subseteq\{1,\dots,t\}$ that satisfies the
following two properties: (1) decisions for nonactive hypotheses are
not updated ($\{\mathcal{R}_{t}\setminus\mathcal{C}_{t}\}=\{\mathcal{R}_{t-1}\setminus\mathcal{C}_{t}\}$),
and (2) for all $i\in\mathcal{R}_{t}$, we have $P_{i}\leq\beta(1\vee|\mathcal{R}_{t}|)A_{i}\alpha$.
The second property mimics the LOND condition (Eq (\ref{eq:lond})),
and will be used to show FDR control. We achieve these two properties
as follows. 

\begin{algorithm}

\label{alg:(TOAD)-Initialize-.}(TOAD) Take as input a function $\beta$
(either the identity function or a shape function), and a value for
$A_{1}$.
\begin{enumerate}
\item (Initialize) Set $\mathcal{R}_{0}=\emptyset$. For any $i\in\mathbb{N}$
such that $\tau_{i}=0$, determine the value for $A_{i}$.
\item For each stage $t$:
\begin{enumerate}
\item (Save past rejections) Define $\mathcal{R}_{t}^{\text{old}}=\mathcal{R}_{t-1}\setminus\mathcal{C}_{t}$
to be the set of previously rejected indices that are no longer being
actively updated. \label{enu:(Enumerate-past-rejections)}
\item (Order test statistics) Let $W_{i}=P_{i}/A_{i}$, and let $W_{(j,t)}$
be the $j^{th}$ lowest value from the set $\{W_{i}\}_{i\in\mathcal{C}_{t}}$,
such that $W_{(1,t)}\leq\dots\leq W_{(\vert\mathcal{C}_{t}\vert,t)}$. 
\item (Define current rejections) Reject the set of indices $\mathcal{R}_{t}=\mathcal{R}_{t}^{\text{old}}\cup\{i\in\mathcal{C}_{t}:W_{i}\leq W_{(S_{t},t)}\}$,
where 
\begin{equation}
S_{t}=\max\{j\leq|\mathcal{C}_{t}|:W_{(j,t)}\leq\alpha\beta(j+|\mathcal{R}_{t}^{\text{old}}|)\}.\label{eq:s-def}
\end{equation}
\item (Set threshold parameters) For any $i>t$ such that $\tau_{i}=t$,
determine the value for $A_{i}$.
\end{enumerate}
\end{enumerate}
\end{algorithm}

While TOAD can retroactively \emph{reject} certain hypotheses, we
show in the supplementary materials that TOAD never \emph{reverses}
a previous rejection (i.e., $\mathcal{R}_{t}\subseteq\mathcal{R}_{t'}$
for any $t<t'$). This monotonicity property is not strictly required
by our framing, but may facilitate the procedure's implementation.
For example, the property can prove useful if it is logistically straightforward
to announce a new discovery, but difficult to retract a previously
announced discovery.

We can think of TOAD as a generalization of both LOND and BH. In
the special case where all rejection decisions must be finalized immediately
(i.e., $\mathcal{C}_{t}=\{t\}$), our procedure reduces to a version
of LOND with dynamically defined threshold parameters. At the other
extreme, if our hypothesis sequence contains a finite number of elements
(denoted by $t_{\text{max}}$), and if all hypotheses remain active
for the entire sequence (i.e., $\mathcal{C}_{t_{\text{max}}}=\{1,\dots,t_{\text{\ensuremath{\max}}}\})$,
then we can recover the BH algorithm setting $A_{i}=1/t_{\text{max}}$
for all $i$, setting $\beta$ to be the identity function, and applying
TOAD at stage $t_{\text{max}}$. 

As an intermediate setting, if hypotheses remain active according
to a block structure then we can recover a procedure that closely
resembles the $\text{Batch}_{\text{BH}}^{\text{}}$ and $\text{Batch}_{\text{BH}}^{\text{PRDS}}$
algorithms described by \citet{Zrnic2020-xl}. In fact, $\text{Batch}_{\text{BH}}^{\text{PRDS}}$
can \emph{also} be seen as a generalization of both BH and LOND \citep{Zrnic2020-xl}.
However, we show in the supplementary materials that any hypothesis
rejected by $\text{Batch}_{\text{BH}}^{\text{PRDS}}$ is also rejected
by TOAD. Our simulations in Section \ref{sec:Simulations} show that
the reverse is not true, and that TOAD typically achieves higher power
than $\text{Batch}_{\text{BH}}^{\text{PRDS}}$.

\subsection{FDR Control \label{subsec:FDR-control}}

Next, we outline sufficient conditions for FDR control. Our first
assumption places restrictions on how the thresholds can be selected.
This assumption can be ensured by design.

\begin{assumption}

\label{assu:deterministic}(Threshold selection) For each $i\in\mathbb{N}$,
$A_{i}$ is a deterministic function of the first $\tau_{i}$ p-values,
denoted by $\mathcal{P}_{\tau_{i}}=\{P_{1},\dots,P_{\tau_{i}}\}$.

\end{assumption}

We consider variations on this assumption in Section \ref{subsec:Examples},
below, in order to allow online behaviors such as adaptive hypothesis
reordering.

Next, we assume that users have access to conditionally valid test
statistics for each hypothesis. Specifically, we assume that the p-value
$P_{i}$ for any true null $H_{i}$ is conditionally (super)uniformly
distributed, given the information used to select $A_{i}$. 

\begin{assumption}

\label{assu:super-unif}(Conditional super-uniformity) For any $i\in\mathcal{H}_{0}$,
 we have $\mathbb{P}(P_{i}\leq u|\mathcal{P}_{\tau_{i}})\leq u$
for all $u\in[0,1]$ and all realizations of $\mathcal{P}_{\tau_{i}}$.

\end{assumption}

This assumption is based on super-uniformity assumptions used by \citeauthor{Foster2008-ta}
(\citeyear{Foster2008-ta}, see their Eq (10)); \citeauthor{Aharoni2014-lv}
(\citeyear{Aharoni2014-lv}, see their Assumption 1); \citet{Ramdas2017-rm,Xu2020-qx}
and \citet{Zrnic2021-jj}. The assumption is also conceptually similar
to a condition used by \citeauthor{Javanmard2015-tn} (\citeyear{Javanmard2015-tn},
see their Eq (8)). 

Assumption \ref{assu:super-unif} also highlights the benefits of
selecting parameters $A_{i}$ in advance of when they are used (i.e.,
setting $\tau_{i}<i-1$). As \citet{Zrnic2021-jj} point out, the
further we plan in advance, the fewer dependencies we will need to
account for when specifying p-values that satisfy Assumption \ref{assu:super-unif}.
\citeauthor{Zrnic2021-jj} also note that setting parameters in advance
is a natural way to capture the logistical delays that can occur between
test specification and test completion.

Next, we define a condition regarding positive dependence of the p-values.

\begin{assumption}

\label{assu:(Conditional-positive-dependence}(Conditional positive
dependence) For any set of positive integers $\{t,r,i\}$ satisfying
$r,i\leq t$ and $H_{i}\in\mathcal{H}_{0}$, the probability 
\[
\mathbb{P}(1\vee\vert\mathcal{R}_{t}\vert\leq r|P_{i}\leq u,\mathcal{P}_{\tau_{i}})
\]
 is nondecreasing in $u$.

\end{assumption}

Roughly speaking, Assumption \ref{assu:(Conditional-positive-dependence}
says that higher p-values imply a higher probability that $\vert\mathcal{R}_{t}\vert$
is small. The supplementary materials explore this assumption in more
detail, and discuss a connection to the conventional assumption of
``positive regression dependence on a subset'' (PRDS; \citealp{Benjamini2001-ps}). 

We are now prepared to show FDR control for our procedure.
\begin{thm}
\label{thm:(LOAD-FDR-control-PRDS}(FDR Control) Under Assumptions
\ref{assu:deterministic} \& \ref{assu:super-unif}, TOAD satisfies
$\text{FDR}(t)\leq\alpha$ for any $t\in\mathbb{N}$ if either of
the following conditions hold:
\begin{enumerate}
\item \label{enu:(Positive-dependence)-If}(Positive dependence) Assumption
\ref{assu:(Conditional-positive-dependence} holds and $\beta$ is
the identity function; or
\item \label{enu:(General-dependence)-If}(General dependence) $\beta$
is a shape functions in the form of Eq (\ref{eq:shape}).
\end{enumerate}
\end{thm}
If our hypothesis sequence has a finite length, then a natural consequence
of the above result is that $\mathbb{E}\left[\text{FDR}(T)\right]$
is also controlled for random, adaptively determined stopping times
$T$. Because our parameters $A_{t}$ are already adaptively determined,
we can incorporate an adaptive stopping time $T$ by simply setting
$A_{t}=0$ for all $t>T$, and completing the test procedure up to
and including the final stage. 

That said, there are two important caveats to this way of capturing
adaptive stopping times. The first is that certain adaptive stopping
rules may lead to violations of Assumption \ref{assu:(Conditional-positive-dependence},
requiring us to either carefully verify this assumption or to appeal
to Part \ref{enu:(General-dependence)-If} of Theorem \ref{thm:(LOAD-FDR-control-PRDS}
instead. The second is that these forms of adaptive stopping rules
become limited when researchers set parameters $A_{i}$ several stages
in advance ($\tau_{i}<i-1$). By specifying the parameter for a future
test, a researcher also implicitly commits to \emph{completing} that
future test. Although they can adaptively choose to stop all testing
for stages where parameters have not yet been determined, they cannot
choose to avoid tests that have already been specified.

\section{SIMULATIONS\label{sec:Simulations}}

In this section, we investigate the effect of the deadline structure
on TOAD's power. We also compare TOAD against two methods introduced
by \citet{Zrnic2020-xl}, and against a ``naive'' version of BH. 

We adopt a simulation setup based the one used by \citeauthor{Zrnic2020-xl}
(\citeyear{Zrnic2020-xl}; differences are noted below). We define
a sequence of $t_{\text{max}}=3000$ test statistics $(Z_{1},\dots Z_{t_{\text{max}}})\sim N(\mu,\Sigma)$,
where $\mu=(\mu_{1},\dots,\mu_{t_{\text{max}}})$ is a sequence of
mean parameters and $\Sigma$ is a covariance matrix defined in detail
below. For each test statistic $Z_{i}$, our null hypothesis $H_{i}$
is that $E(Z_{i})=0$, and our alternative hypothesis is that $E(Z_{i})=3$.
We use $\pi_{1}$ to denote the proportion of null hypotheses that
are false. In each simulation iteration, we select a random subset
of $\lceil(1-\pi_{1})t_{\max}\rceil$ indices for which we set $\mu_{i}=0$
(i.e., we simulate $Z_{i}$ from the null distribution). We set the
remaining mean parameters equal to 3. 

To define deadline parameters, we will say that hypotheses remain
active within ``batches'' of tests, and use $n_{\text{batch}}$
to denote the batch size. For each $i\in\{1,\dots,t_{\text{max}}\}$,
we set the deadline $d_{i}$ to be the smallest multiple of $n_{\text{batch}}$
that is no less than $i$, that is, $d_{i}=\min\{kn_{\text{batch}}:k\in\mathbb{N}\text{ and }i\leq kn_{\text{batch}}\}$.
For example, if $n_{\text{batch}}=100,$ then $d_{i}=100$ for $i\in[1,100]$;
$d_{i}=200$ for $i\in[101,200]$; and so on. We define $\Sigma$
so that $Var(Z_{i})=1$ for all $i$; $Cov(Z_{i},Z_{j})=\rho$ if
$i\neq j$, but $i$ and $j$ are in the same batch; and $Cov(Z_{i},Z_{j})=0$
if $i$ and $j$ are not in the same batch

We simulate all combinations of $\rho\in\{0,0.5\}$; $n_{\text{batch}}\in\{10,100,1000\}$;
and 
\[
\pi_{1}\in\{0.01,0.02,\dots,0.09,0.1,0.2,0.3,0.4,0.5\}.
\]
For each combination, we simulate $500$ iterations. 

Our simulation setup differs from that of \citet{Zrnic2020-xl} in
two ways. Most notably, \citeauthor{Zrnic2020-xl} only simulate the
case where $\rho=0$, as most of the methods they develop are designed
for the case of independent test statistics. \citeauthor{Zrnic2020-xl}
also use a Bernoulli distribution to determine whether each test statistic
$Z_{i}$ is generated from a null distribution or an alternative distribution,
meaning that the realized proportion of truly null hypotheses varies
slightly across simulation iterations.

\subsection{Comparator Methods}

As comparators for TOAD, we primarily consider the $\text{Batch}_{\text{BH}}$
and $\text{Batch}_{\text{BH}}^{\text{PRDS}}$ algorithms \citep{Zrnic2020-xl}.
The first method, $\text{Batch}_{\text{BH}}$, is proven to control
FDR under an independence assumption. The second method, $\text{Batch}_{\text{BH}}^{\text{PRDS}}$,
is proven to control FDR if test statistics are independent across
batches and positively dependent within each batch. Thus, we expect
$\text{Batch}_{\text{BH}}$ to achieve higher power than $\text{Batch}_{\text{BH}}^{\text{PRDS}}$,
potentially at the cost of FDR control. 

For the tuning parameters of TOAD, we set $\beta$ equal to the identity
function, and set $\tau_{i}=0$ and $A_{i}=1/t_{\text{max}}$ for
all $i$. Similarly, for $\text{Batch}_{\text{BH}}^{\text{PRDS}}$,
we use the implementation defined in \citeauthor{Zrnic2020-xl}'s
appendix, and use tuning parameters that place equal weight on each
batch. For $\text{Batch}_{\text{BH}}$, we use the implementation
and tuning parameters described in \citeauthor{Zrnic2020-xl}'s simulations. 

We also compare against the ``naive'' approach of running BH separately
in each batch at an alpha level of $\alpha(t_{\text{max}}/n_{\text{batch}})^{-1}$,
where $t_{\text{max}}/n_{\text{batch}}$ is the number of batches.
We refer to this last method as ``Naive-BH.'' For completeness,
we briefly show in the supplementary materials that Naive-BH also
controls the false discovery rate whenever the p-values are positively
dependent.

For all of the above methods, we set $P_{i}=\Phi(-Z_{i})$, where
$\Phi$ is the CDF of a standard normal distribution. That is, we
define each p-value to be the result of a one-sided test of $H_{i}$.

\subsection{Simulation Results}

\begin{figure}
\centering{}\includegraphics[width=0.6\columnwidth]{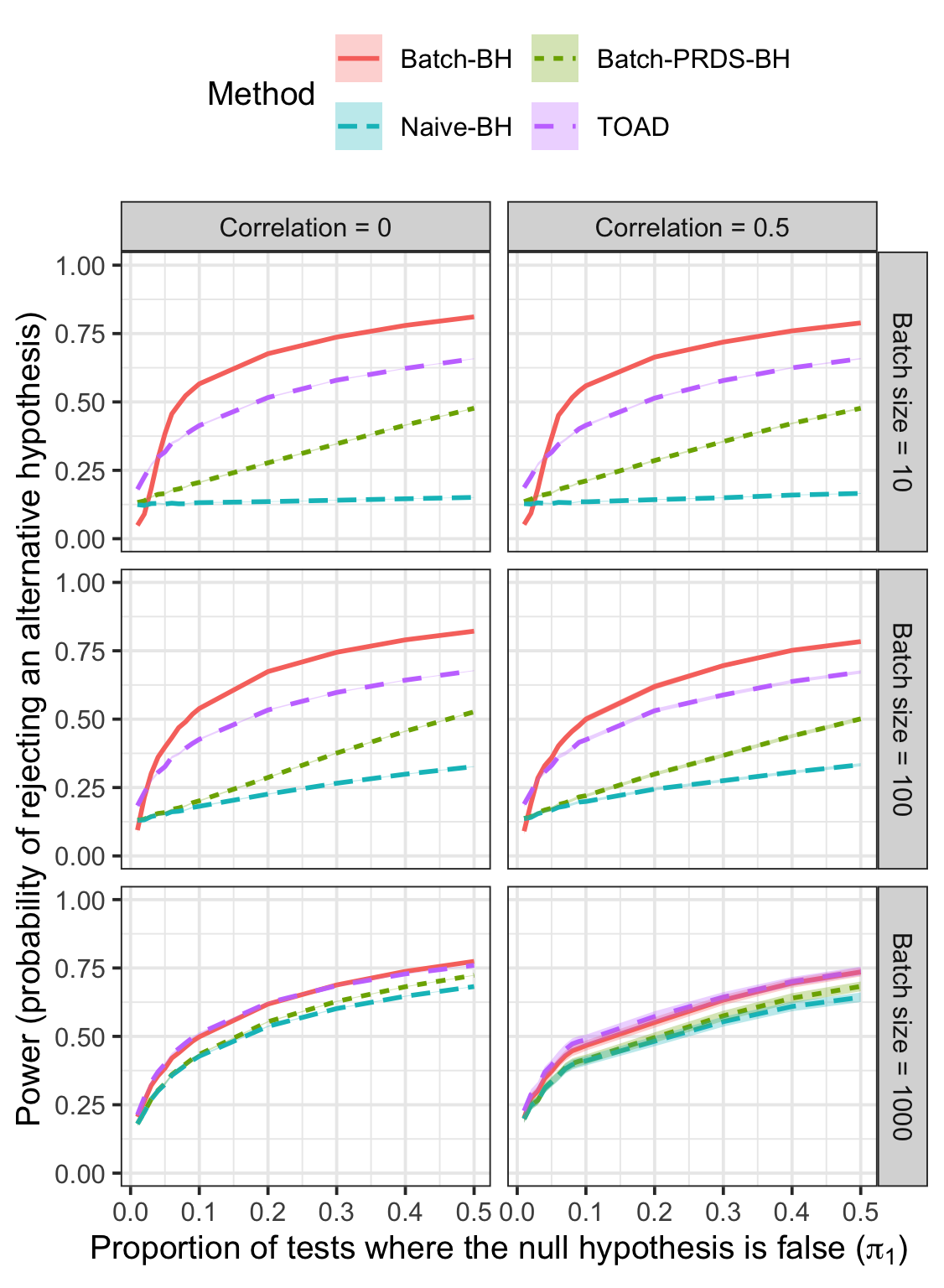}\caption{\label{fig:Simulated-power-for-1}Simulated power for each method
-- We simulate test statistics under a ``batch'' structure, where
all hypotheses in a batch share a common deadline. The test statistics
are normally distributed with possible within-batch correlation (denoted
by columns). For each null hypothesis $H_{i}:\mathbb{E}(Z_{i})=0$,
we generate one-sided p-values as $\Phi(-Z_{i})$, where $\Phi$ is
the cumulative distribution function for a standard normal distribution.
Shaded ribbons show a range of $\pm$ two Monte Carlo standard errors
($\sqrt{\frac{1}{500}\text{Var}\left(\vert\mathcal{R}_{t_{\text{max}}}\cap\bar{\mathcal{H}}_{0}\vert/\vert\bar{\mathcal{H}}_{0}\vert\right)}$,
where $500$ is the number of simulation iterations and $\bar{\mathcal{H}}_{0}$
is the set of false nulls), although these errors are negligible in
many cases. The $\text{Batch}_{\text{BH}}$ method generates the highest
power, but also requires the strongest assumptions in order to guarantee
control of the FDR. Of the methods that ensure FDR control for positively
dependent test statistics, TOAD achieves the highest power.}
\end{figure}

\begin{figure}
\centering{}\includegraphics[width=0.6\columnwidth]{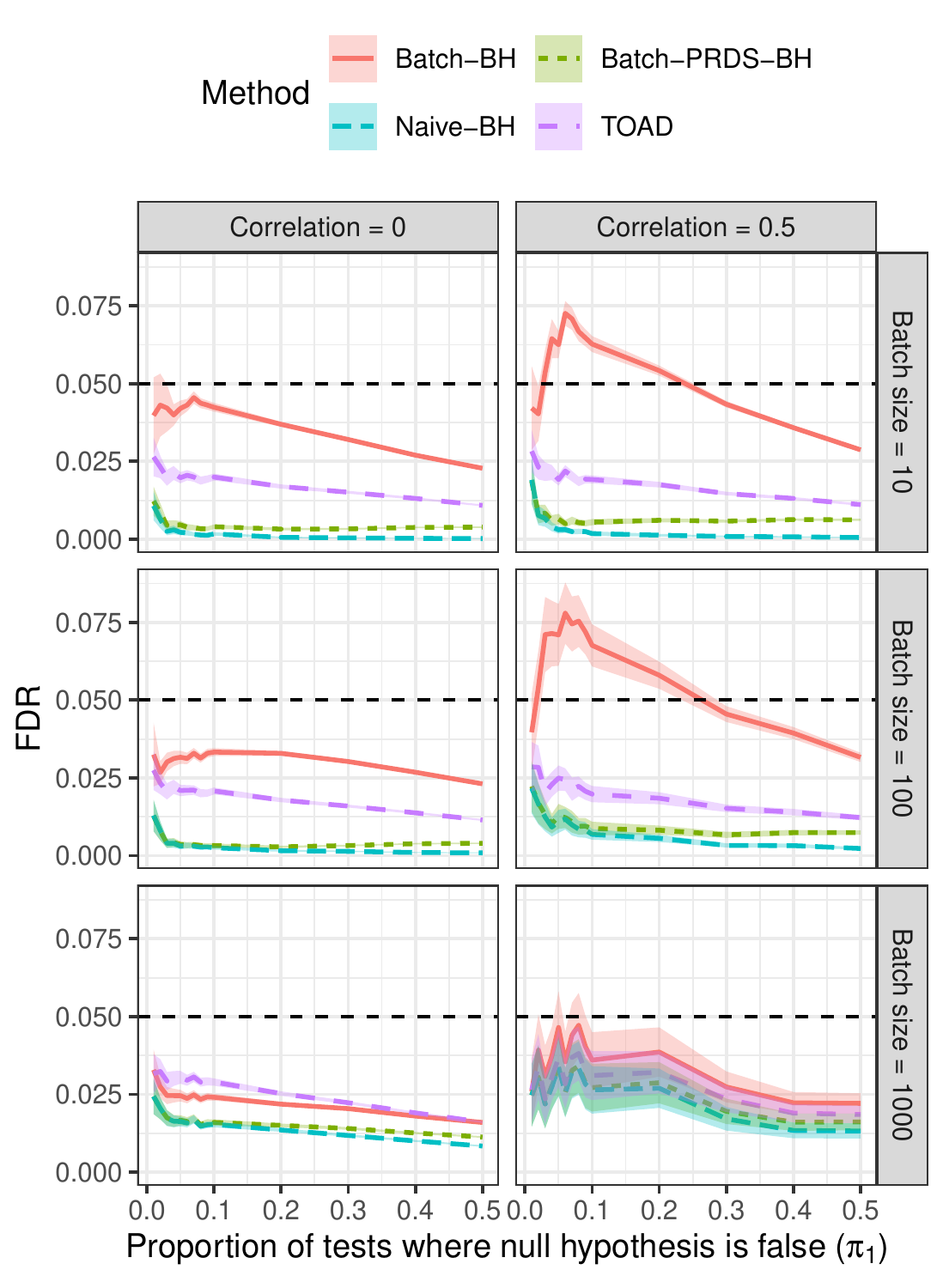}\caption{\label{fig:Simulated-fdr-cor-5}Simulated FDR for each method --
Again, shaded ribbons show a range of $\pm$ two Monte Carlo standard
errors ($\sqrt{\frac{1}{500}\text{Var}\left(\vert\mathcal{H}_{0}\cap\mathcal{R}_{t_{\text{max}}}\vert/\left(1\vee\vert\mathcal{R}_{t_{\text{max}}}\vert\right)\right)}$,
where $500$ is the number of simulation iterations). The dashed line
shows our desired FDR level. We see that the power of $\text{Batch}_{\text{BH}}$
can come at the cost of inflated FDR in the face of within-batch correlation
(right column). }
\end{figure}

Figure \ref{fig:Simulated-power-for-1} shows the simulated power
for each method tested, where power is defined as the expected proportion
of alternative hypotheses that are rejected in any one experiment.
\textcolor{black}{Figure \ref{fig:Simulated-fdr-cor-5}} shows the
FDR for each procedure. 

$\text{Batch}_{\text{BH}}$ consistently generates the highest power,
with TOAD generating the second highest. The one exception comes when
batches sizes are large ($b=1000$), in which case TOAD and $\text{Batch}_{\text{BH}}$
have comparable power.\textcolor{black}{{} To some extent, this is to
be expected, as TOAD provides stronger FDR guarantees than $\text{Batch}_{\text{BH}}$
does. Indeed, we see that when the assumptions of }$\text{Batch}_{\text{BH}}$
are violated due to within-batch correlation, \textcolor{black}{$\text{Batch}_{\text{BH}}$
produced an inflated FDR (see Figure \ref{fig:Simulated-fdr-cor-5}).}

On the other hand, \textcolor{black}{$\text{Batch}_{\text{BH}}^{\text{PRDS}}$
offers FDR guarantees that are more comparable to those of TOAD. Thus,
$\text{Batch}_{\text{BH}}^{\text{PRDS}}$ forms an especially informative
comparator. We see that TOAD has higher power than $\text{Batch}_{\text{BH}}^{\text{PRDS}}$
across all scenarios, as we would expect from our analytical result
in }the supplementary materials\textcolor{black}{. }

In addition to these simulations, we also considered the setup described
by \citet{Zrnic2020-xl} in which each mean parameter $\mu_{i}$ corresponding
to the alternative distribution is randomly generated. This results
in some test statistics carrying strong signal while others carry
only weak signal. Similar patterns occurred in this setting, although
the differences between all four methods were less pronounced (see
details in the supplemental materials).

\section{DISCUSSION}

We have proposed an online version of the \citet{Benjamini1995-sz}
method that includes limited forms of decision updating. Our procedure
controls the FDR under arbitrary p-value dependence structures, and
at adaptively determined stopping times. Compared to similar procedures
with comparable FDR guarantees, we find that our approach also provides
superior power. 

We conclude by discussing several immediate extensions.

\subsection{Ignoring Hypotheses, and Adaptive Hypothesis Reordering\label{subsec:Examples}}

A central advantage of online procedures is their ability to selectively
ignore hypotheses based on preliminary results. Here, we say that
a hypothesis $H_{i}$ is ``ignored'' if $A_{i}=0$ (see also Appendix
B of \citealp{Ramdas2017-rm} for a similar discussion). Using the
idea of ignoring hypotheses as a building block, we can quickly encompass
other types of online strategies. For example, if the hypothesis sequence
$H_{1},H_{2},\dots$ is sufficiently diverse, then we can effectively
\emph{define} our hypotheses adaptively by ignoring those hypotheses
that are no longer of interest. 

Similarly, ignoring hypotheses effectively lets us adaptively\emph{
reorder} the available hypotheses. For example, suppose that a researcher
plans to test three unique hypotheses $\tilde{H}^{(1)},\tilde{H}^{(2)},\tilde{H}^{(3)}$,
but wishes to test the last two in an adaptive order. This can be
achieved by defining the expanded, 5-stage hypothesis sequence 
\[
(H_{1},H_{2},H_{3},H_{4},H_{5})=(\tilde{H}^{(1)},\tilde{H}^{(2)},\tilde{H}^{(3)},\tilde{H}^{(2)},\tilde{H}^{(3)}),
\]
 shown in Table \ref{tab:Adaptive-hypothesis-reordering}. From here,
depending on how the parameters $(A_{2},A_{3},A_{4},A_{5})$ are selected,
the researcher can use the result of the first test to decide whether
to test $\tilde{H}^{(2)}$ before $\tilde{H}^{(3)}$, or vice versa
(see details in Table \ref{tab:Adaptive-hypothesis-reordering}).
The same approach can be used to reorder arbitrarily large hypothesis
sets. 
\begin{table}
\caption{\label{tab:Adaptive-hypothesis-reordering}Online Hypothesis Reordering}

\medskip{}

\begin{centering}
\begin{tabular}{cccc}
\textbf{STAGE } & \textbf{$H_{t}$ } & \textbf{OPTION 1 } & \textbf{OPTION 2}\tabularnewline
\textbf{($t$)} &  & \textbf{FOR $A_{t}$} & \textbf{FOR $A_{t}$}\tabularnewline
\hline 
1 & $\tilde{H}^{(1)}$ & 1/3 & 1/3\tabularnewline
2 & $\tilde{H}^{(2)}$ & 1/3 & 0\tabularnewline
3 & $\tilde{H}^{(3)}$ & 0 & 1/3\tabularnewline
4 & $\tilde{H}^{(2)}$ & 0 & 1/3\tabularnewline
5 & $\tilde{H}^{(3)}$ & 1/3 & 0\tabularnewline
\end{tabular}
\par\end{centering}
\medskip{}

\medskip{}

\begin{raggedright}
Table \ref*{tab:Adaptive-hypothesis-reordering} Caption: The first
column shows the stage index for a 5-stage experiment. The second
column shows a sequence of hypotheses, including duplicates, to be
tested in an online fashion at each stage. The third and fourth columns
offer different choices for the tuning parameters $A_{2},\dots,A_{5}$,
where the choice between these options can be made at the end of Stage
1 (i.e., after observing $P_{1}$). Option 1 amounts to testing the
hypotheses in the order $\tilde{H}^{(1)},\tilde{H}^{(2)},\tilde{H}^{(3)}$,
while Option 2 amounts to testing the hypotheses in the order $\tilde{H}^{(1)},\tilde{H}^{(3)},\tilde{H}^{(2)}.$
\par\end{raggedright}
\medskip{}

\medskip{}
\end{table}

In order to leverage the benefits of ignoring hypotheses, we will
need restrict the information used to define upcoming threshold parameters
$A_{i}$. At present, our Assumption \ref{sec:Thresholds-Based-on}
requires that future test statistics be conditionally uniform given
the previous p-values, and such a condition can be impossible to satisfy
if the hypothesis sequence contains repeats. For this reason, we suggest
modifying Assumptions \ref{assu:deterministic}, \ref{assu:super-unif}
\& \ref{assu:(Conditional-positive-dependence} so that testing decisions
depend only on the previous ``unignored'' hypotheses. To formalize
this, we define $P_{t}^{\text{obs}}=P_{t}\times1(A_{t}>0)-1(A_{t}=0)$
to be equal to $-1$ if $H_{t}$ is ignored and equal to $P_{t}$
otherwise. Thus, the sequence $\mathcal{P}_{\tau_{i}}^{\text{obs}}=\{P_{i'}^{\text{obs}}\}_{i'\leq\tau_{i}}$
contains the information in the first $\tau_{i}$ p-values that is
not ignored. Our Theorem \ref{thm:(LOAD-FDR-control-PRDS} is unchanged
if we replace $\mathcal{P}_{\tau_{i}}$ with $\mathcal{P}_{\tau_{i}}^{\text{obs}}$
in Assumptions \ref{assu:deterministic}, \ref{assu:super-unif} \&
\ref{assu:(Conditional-positive-dependence} (see the proof of Theorem
\ref{thm:(LOAD-FDR-control-PRDS} in the supplementary materials). 

\subsection{Forgetting Antiquated Results\label{subsec:Forgetting-antequated-results}}

\citet{Ramdas2017-rm} remark that, in short-term forecasting problems,
hypotheses tested in the distant past have little bearing on our decisions
at present. With this in mind, they propose a ``decaying memory''
variation of FDR that places more weight on recently tested hypotheses.
That is, they focus on multiplicity corrections for the discoveries
currently in use, rather than for all discoveries made over the course
of an experiment. 

In some ways, the idea that hypotheses from the distant past carry
less importance at present is a natural complement to the idea that
hypotheses eventually pass a deadline beyond which any retroactive
discovery is irrelevant. Thus, one fruitful avenue of future research
could be to formally blend the ideas of decaying memory and deadlines. 

A simply way of doing this is to omit ``outdated'' or ``forgotten''
hypotheses from the FDR computation, resulting in
\[
\text{FDR}_{\text{recent}}(t)=\mathbb{E}\left[\frac{\vert\mathcal{H}_{0}\cap\mathcal{R}_{t}\cap\mathcal{C}_{t}\vert}{1\vee\vert\mathcal{R}_{t}\cap\mathcal{C}_{t}\vert}\right].
\]
It is straightforward to show that TOAD controls $\text{FDR}_{\text{\text{recent}}}(t)$
if we (1) relax the requirement that $\sum_{i=1}^{\infty}A_{i}\leq1$
to instead require that $\sum_{i\in\mathcal{C}_{t}}A_{i}\leq1$ for
all $t$, and (2) replace $\mathcal{R}_{t}^{\text{old}}$ with the
empty set $\emptyset$ throughout the procedure (see the supplementary
materials). Under such a procedure, the parameters $A_{i}$ from outdated
hypotheses can be ``recycled'' towards future tests. 

However, an important caveat is that $\text{FDR}_{\text{\text{recent}}}(T)$
is more difficult to control under adaptive stopping times $T$. Before,
we were able to control $\text{FDR}(T)$ simply by controlling $\text{FDR}(t_{\text{max}})$
(Section \ref{subsec:FDR-control}). Here though, controlling $\text{FDR}_{\text{\text{recent}}}(t_{\text{max}})$
is not sufficient for controlling $\text{FDR}_{\text{\text{recent}}}(T)$.
Roughly speaking, $\text{FDR}_{\text{\text{recent}}}(t_{\text{max}})$
``forgets'' the information that would have been necessary to control
error rates at earlier times.

\section*{ACKNOWLEDGEMENTS}

The author is grateful for a helpful correspondence with Tijana Zrnic while developing this work.

\bibliographystyle{apalike}
\bibliography{_repeated-analysis}

\end{document}